\documentclass{emulateapj} 
\usepackage{apjfonts} 
\pdfoutput=1
\usepackage{epsfig}

\newcommand{\arcm}{\hbox{$^\prime$}}

\newcommand{\degree}{\hbox{$^\circ$}}

\newcommand{\chandra}{\emph{Chandra}}
\newcommand{\xmm}{\emph{XMM-Newton}}
\newcommand{\xmms}{\emph{XMM}}
\newcommand{\asca}{\emph{ASCA}}

\newcommand{\swift}{\emph{Swift}}
\newcommand{\arcs}{\mbox{\arcm\arcm}}

\newcommand{\Zsol}{\ensuremath{~Z_{\odot}}}

\newcommand{\Msol}{\ensuremath{~M_{\odot}}}

\newcommand{\s}{\ensuremath{\mbox{~s}}}
\newcommand{\ps}{\ensuremath{\s^{-1}}}
\newcommand{\cm}{\ensuremath{\mbox{~cm}}}
\newcommand{\pcmsq}{\ensuremath{\cm^{-2}}}
\newcommand{\pcmcu}{\ensuremath{\cm^{-3}}}
\newcommand{\km}{\ensuremath{\mbox{~km}}}

\newcommand{\erg}{\ensuremath{\mbox{~erg}}}
\newcommand{\ergps}{\ensuremath{\erg \ps}}

\newcommand{\kmps}{\ensuremath{\km \ps}}

\newcommand{\Hi}{H\textsc{i}}
\newcommand{\Ha}{\ensuremath{\mathrm{H\alpha}}}
\newcommand{\nh}{\ensuremath{\mathrm{n}_\mathrm{H}}}

\begin{document}

\title{A Chandra X-ray view of Stephan's Quintet: Shocks and Star-formation}

\author{E. O'Sullivan\altaffilmark{1}, S. Giacintucci\altaffilmark{1}, J.
  M. Vrtilek\altaffilmark{1}, S. Raychaudhury\altaffilmark{2} and L. P.
  David\altaffilmark{1}} \altaffiltext{1}{Harvard-Smithsonian Center for
  Astrophysics, 60 Garden Street, Cambridge, MA 02138, USA, email:
  \textit{eosullivan@cfa.harvard.edu}.} \altaffiltext{2}{School of Physics
  and Astronomy, University of Birmingham, Edgbaston, Birmingham B15 2TT,
  UK.} 
\shorttitle{A Chandra view of Stephan's Quintet}
\shortauthors{O'Sullivan et al}

\begin{abstract}
  We use a deep \chandra\ observation to examine the structure of the hot
  intra--group medium of the compact group of galaxies Stephan's Quintet.
  The group is thought to be undergoing a strong dynamical interaction as
  an interloper, NGC~7318b, passes through the group core at
  $\sim850$\kmps. Previous studies have interpreted a bright ridge of
  X--ray and radio continuum emission as the result of shock heating, with
  support from observations at other wavelengths. We find that gas in this
  ridge has a similar temperature ($\sim0.6$~keV) and abundance
  ($\sim0.3$\Zsol) to the surrounding diffuse emission, and that a hard
  emission component is consistent with that expected from high--mass
  X--ray binaries associated with star--formation in the ridge. The cooling
  rate of gas in the ridge is consistent with the current star formation
  rate, suggesting that radiative cooling is driving the observed star
  formation.  The lack of a high--temperature gas component is used to
  place constraints on the nature of the interaction and shock, and we find
  that an oblique shock heating a pre--existing filament of \Hi\ may be the
  most likely explanation of the X--ray gas in the ridge. The mass of hot
  gas in the group is roughly equal to the deficit in observed \Hi\ mass
  compared to predictions, but only $\sim2\%$ of the gas is contained in
  the ridge. The hot gas component is too extended to have been heated by
  the current interaction, strongly suggesting that it must have been
  heated during previous dynamical encounters.
\end{abstract}

\keywords{galaxies: clusters: individual (Stephan's Quintet; HCG 92) --- galaxies: intergalactic medium --- galaxies: interactions --- X-rays: galaxies}

\section{Introduction}
Stephan's Quintet, also known as \object[HCG92]{HCG~92} \citep[][hereafter
SQ]{Hickson82}, was discovered roughly 130 years ago \citep{Stephan1877}
and is perhaps the most extensively studied compact galaxy group. Several
factors contribute to the interest shown in the system. Of the original
five galaxies identified as part of the group, one
(\object[NGC7320]{NGC~7320}) has a discordant redshift and is now
recognised as a superimposed foreground object; attempts to determine the
true distances of the various members motivated much early work. Among the
remaining galaxies, three (\object[NGC7317]{NGC~7317},
\object[NGC7318a]{NGC~7318a} and \object[NGC7319]{NGC~7319}) form a central
kernel with similar recession velocities, while the fifth
(\object[NGC7318b]{NGC~7318b}) appears to be passing through the group core
with a velocity of $\sim$850\kmps. The group shows signs of complex past
tidal interactions, including stellar and \Hi\ tidal tails
\citep{Arp73,Shostaketal84,Sulenticetal01,Williamsetal02} hosting ongoing
star formation
\citep{Xuetal99,Gallagheretal01,MendesdeOliveiraetal04,Xuetal05}, and
possible tidal dwarf galaxies \citep{Xuetal03}. These features were
probably formed during one or more passages through the group by the nearby
\object[NGC7320c]{NGC~7320c} \citep{Shostaketal84,Molesetal97}.

However, perhaps the most interesting aspect is the evidence of ongoing
interactions as NGC7318b passes through the group. 1.4~GHz radio continuum
observations reveal a ridge of emission lying along the eastern edge of
NGC~7318b \citep{AllenHartsuiker72,VanderhulstRots81} which has been shown
to correspond to similar features in \Ha\ \citep{Sulenticetal01}, UV
\citep{Xuetal05}, H$_2$ \citep{Appletonetal06}, 15$\mu$m IR
\citep{Xuetal99} and X-ray \citep{Pietschetal97} emission. This structure
has been widely interpreted as a shock caused by the collision of NGC~7318b
with a pre-existing IGM or tidally stripped material
\citep[e.g.,][]{Sulenticetal01,Xuetal03}. As such, it may serve as an
example of the process by which the X-ray emitting IGM seen in more evolved
groups is formed.

SQ has been observed previously by both \chandra\ and \xmm\
\citep{Trinchierietal03,Trinchierietal05}. Unfortunately the IGM is not
particularly bright in X-rays ($\sim$2$\times$10$^{41}$\ergps\ for the
shock/ridge region), the \chandra\ pointing was relatively short ($<$20~ks)
and the group is quite compact compared to the \xmms\ field of view and
imaging resolution. Spectral fitting of the ridge suggested a
multi-temperature model was needed but did not strongly constrain the shock
properties. The data did show the presence of substructures within the
ridge, and indicated the possibility of temperature differences between
these structures.

In this paper we combine a new, deeper \chandra\ exposure with the
previous short observation to provide stronger constraints on the
properties of the ridge and associated shock, and the surrounding IGM. SQ
is an exceptionally complex system, with X-ray emission arising from many
other sources, including the individual galaxies, tidal features and star
forming regions.  We will discuss these in detail and compare them with
results from our own observations in other wavebands, which will in turn be described in detail in future papers.

\section{Observations and Data Analysis}
SQ was first observed by the \chandra\ ACIS instrument during Cycle~1 on
2000 July 09 (\dataset[ADS/Sa.CXO#obs/789]{ObsId 789}), for $\sim$20~ks,
and then again during Cycle~8, on 2007 August 17-18
(\dataset[ADS/Sa.CXO#obs/7924]{ObsId 7924}) for $\sim$95~ks. A summary of
the \chandra\ mission and instrumentation can be found in
\citet{Weisskopfetal02}. In both cases the S3 CCD was placed at the focus
of the telescope. For the first observation the instrument operated in
faint mode, while the second was performed in very faint mode to take
advantage of the superior background cleaning. We have reduced the data
from both pointings using CIAO 4.0.1 and CALDB 3.4.5 following techniques
similar to those described in \citet{OSullivanetal07} and the \chandra\
analysis threads\footnote{http://asc.harvard.edu/ciao/threads/index.html}.
The level 1 events files were reprocessed, bad pixels and events with
\asca\ grades 1, 5 and 7 were removed, and the cosmic ray afterglow
correction was applied. The data were corrected to the appropriate gain
map, the standard time-dependent gain and charge-transfer inefficiency
(CTI) corrections were made, and a background light curve was produced.
Neither observation suffered from significant background flaring, and the
final cleaned exposure times were 19.7 and 93.2~ks respectively. In general
the observations were combined for imaging analysis, but spectra were
extracted from the longer observation only.

Point source identification was performed using the \textsc{ciao} task
\textsc{wavdetect}, with a detection threshold of 10$^{-6}$, chosen to
ensure that the task detects $\leq$1 false source in the field, working
from a 0.3-7.0 keV image and exposure map from the combined observations.
Source ellipses were generated with axes of length 4 times the standard
deviation of each source distribution. These were then used to exclude
sources from most spectral fits.

Spectra were extracted using the \textsc{specextract} task. Background
spectra were drawn from a rectangular region of size $\sim$120x900 pixels,
running parallel to the detector boundary north of the group emission. The
region was selected to lie $\sim$120 pixels from the CCD edge, so as to
avoid areas where the contaminant on the optical filter could produce
increased absorption at low energies. A local background is preferred when
fitting spectra where energies $<$1~keV are of interest, as it will have
the correct hydrogen column (whereas blank-sky data are assembled from
observations with a range of columns) and will correctly subtract any
contamination from solar wind charge exchange emission. However, to ensure
that our local background region contained no source emission, we
extracted alternate background spectra from the blank-sky events lists,
scaled to match the data in the 9.5-12.0 keV band and corrected by
comparing the source and blank-sky spectra on the source-free S1 chip. We
found no significant difference in our results when using either method,
and conclude that our choice of local background region is satisfactory.
Spectral fitting was performed in XSPEC 11.3.2ag. Abundances were measured
relative to the abundance ratios of \citet{GrevesseSauval98}. A galactic
hydrogen column of 6.17$\times10^{20}$\pcmsq\ was assumed in most fits. 90\%
errors are reported for all fitted values. For calculation of luminosities
and scales, we adopt a distance to the group of 85~Mpc.

\subsection{Temperature Mapping}

To examine the spatial variation of temperature in the gas, we prepared a
temperature map using the technique developed by David et al. (in prep.),
which takes advantage of the close correlation between the strength of
lines in the Fe-L complex and gas temperature in $\sim$1~keV plasma.
Most of the emission from such gas arises from the L--shell lines from
Fe-XIX (Ne--like) to Fe-XXIV (He--like). For CCD resolution spectra, these
lines are blended to form a single broad peak between approximately 0.7 and
1.2 keV. The centroid or mean photon energy of this peak increases with the
temperature of the gas as the dominant ionization state of Fe shifts from
Fe XIX in 0.5~keV gas to Fe XXIV in 1.2 keV gas. Since Li--like Fe is the
highest ionization state that can produce L--shell lines, the mean photon
energy of the blended L--shell lines is independent of energy above
$\sim$1.2~keV.

\begin{figure}[t]
\centerline{\includegraphics[width=8cm,viewport=20 210 558 743]{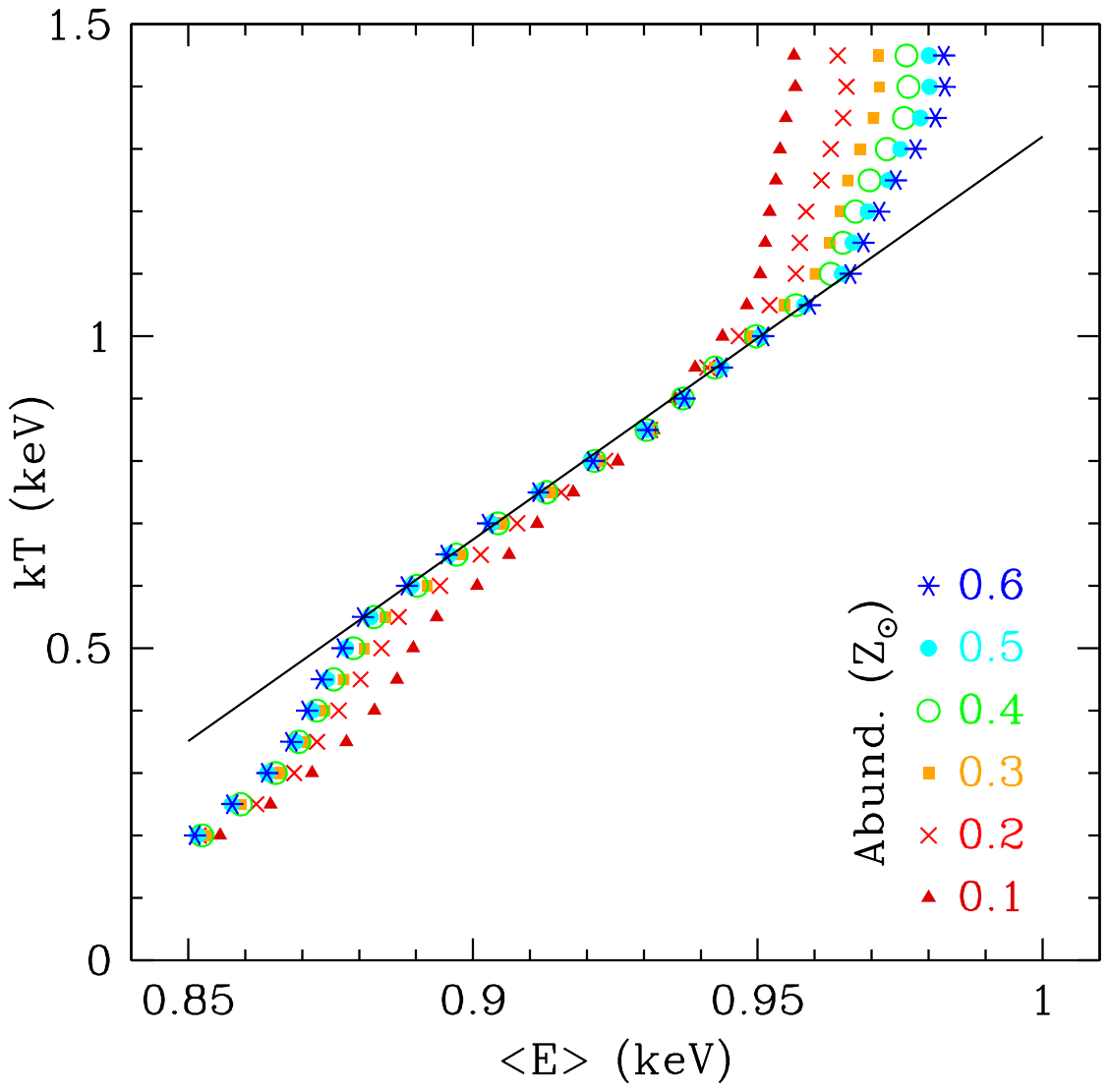}}
\caption{\label{map_fit} Correlation between mean photon energy in the 0.7-1.2 keV energy band and gas temperature as derived from XSPEC simulations. Symbols represent simulations with abundances between 0.1 and 0.6\Zsol, as indicated by the key. The solid line shows the fitted relation used in the temperature map.}
\end{figure}

We can thus estimate the temperature distribution of the gas by mapping the
mean photon energy in the 0.7-1.2~keV band. The relationship between kT and
mean photon energy was determined by simulating spectra based on an
absorbed vapec model with redshift set to that of SQ ($z=0.0215$),
abundance fixed at a typical value determined from spectral fitting
(0.3~\Zsol) and the hydrogen column set to the galactic value. These showed
that for temperatures between $\sim$0.5 and $\sim$1.0~keV, the relationship
is approximately linear ($kT=-5.14+6.46<E>$). We expect the correlation to
be insensitive to global abundance variations, and test this by repeating
the simulations with a range of abundances. Figure~\ref{map_fit} shows that
within our temperature bounds the relation remains reasonably accurate
across a range of abundances. There is significant deviation only at the
lowest abundances (0.1\Zsol) and temperatures ($\sim$0.5~keV), where the
relation may overestimate the true temperature by $\sim$0.1~keV. The
relationship should not be affected by variations in abundance ratio. While
the 0.7-1.2~keV band contains some emission from O, Ne and Mg, at solar
abundance ratios 93\% of the emission arises from Fe, so exceptionally
strong changes of abundance ratio (which would be obvious in spectra) would
be required to affect affect the mean photon energy. We also expect the
correlation to be only mildly sensitive to changes in hydrogen column.
Simulations using a fixed temperature (0.6 keV, typical of SQ) but variable
\nh\ show that changing the column by a factor 2 alters the map temperature
by 0.02-0.04~keV, an acceptable uncertainty for our purposes. The maximum
column due to \Hi\ within SQ has been measured to be
$\sim1.7\times10^{21}$\pcmsq\ \citep{Williamsetal02}; this would increase
the estimated temperature to 0.67~keV, still a relatively small change.

This technique has the advantage of allowing the creation of maps with much
finer resolution, or for regions with fewer detected counts, than would be
possible using spectral fitting. In this case we have adaptively smoothed
the map using the CIAO task \textsc{csmooth}, with smoothing scales based
on the 0.7-1.2~keV image. Comparisons show that temperatures derived from
such maps generally agree well with spectral fits, but we emphasize that we
use the map only as a diagnostic tool, to give a general impression of the
relative temperature structure and help select regions for further
investigation.

\section{Imaging Analysis}
\begin{figure*}[t]
\centerline{\includegraphics[width=\textwidth]{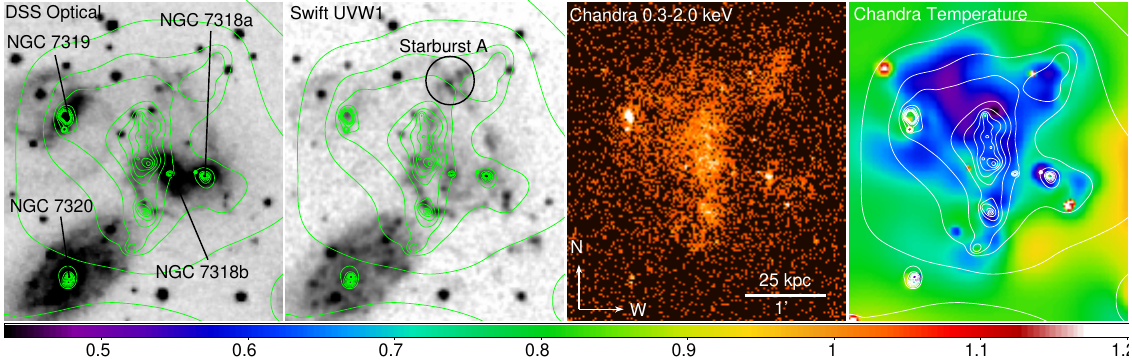}}
\caption{\label{images} From left to right: Digitized Sky Survey POSS2
  $B_j$-band optical image of the core of SQ with galaxies marked; \swift\
  UV/optical telescope UVW1-band image (centered at $\sim$260nm) with the
  ``starburst A'' star forming region identified by \citep{Xuetal99};
  \chandra\ 0.3-2.0~keV exposure corrected X--ray image of the same region,
  using the combined data from both exposures; Adaptively smoothed
  \chandra\ temperature map of SQ (scalebar shows kT in keV). All images
  are on the same scale and have the same alignment. Adaptively smoothed
  X--ray contours spaced by factors of $\sqrt{2}$ are overlaid on panels 1,
  3 \& 4.}
\end{figure*}

Figure~\ref{images} shows images of the core of SQ in the optical, near-UV,
and soft (0.3-2.0~keV) X--ray band, and an X--ray temperature map.  The
images are centered on the emission ridge, which extends north-south along
the eastern edge of NGC~7318b. There is some variation in X--ray brightness
along the ridge, with the northern regions and the southern tip being
brightest, with a fainter region between them. Several X-ray point sources
are also identified in the ridge. The X-ray image shows several other
features reported in previous studies of SQ: point sources associated with
the centres of the four galaxies in the field of view, diffuse emission
extending from the ridge towards NGC~7319, knots of emission northwest of
the ridge and fainter emission coincident with and to the southwest of
NGC~7318a/b.

The temperature map generally agrees with the image in identifying regions
where gas is present. However, the temperature structure does not directly
match the surface brightness structure. This is most notable in two
regions: 1) the coolest gas is found to the northeast of the ridge region,
possibly extending around its northern tip, and 2) the temperature of the
ridge does not appear to be significantly raised above the surrounding
emission, particularly that of the gas between the ridge and NGC~7319. At
its southern end, the ridge is in fact cooler than the surrounding
emission. In general, the ridge is not visible as a distinct structure in
the temperature map.

It is also clear that the temperature is not strongly affected by
absorption effects, as there is little correlation between map features and
the known positions of \Hi\ clouds. The highest hydrogen column in the
group is found around the Starburst A region, and here we see a decrement
in X--ray emission, with brighter regions to the southeast and
northwest. The map temperature is also higher, but only by a marginal
amount ($\sim$0.05~keV). We therefore conclude that the map is a reliable
guide to the temperature structure, particularly in the ridge where
observations show there to be little or no \Hi.

\begin{figure}
\centerline{\includegraphics[width=8cm]{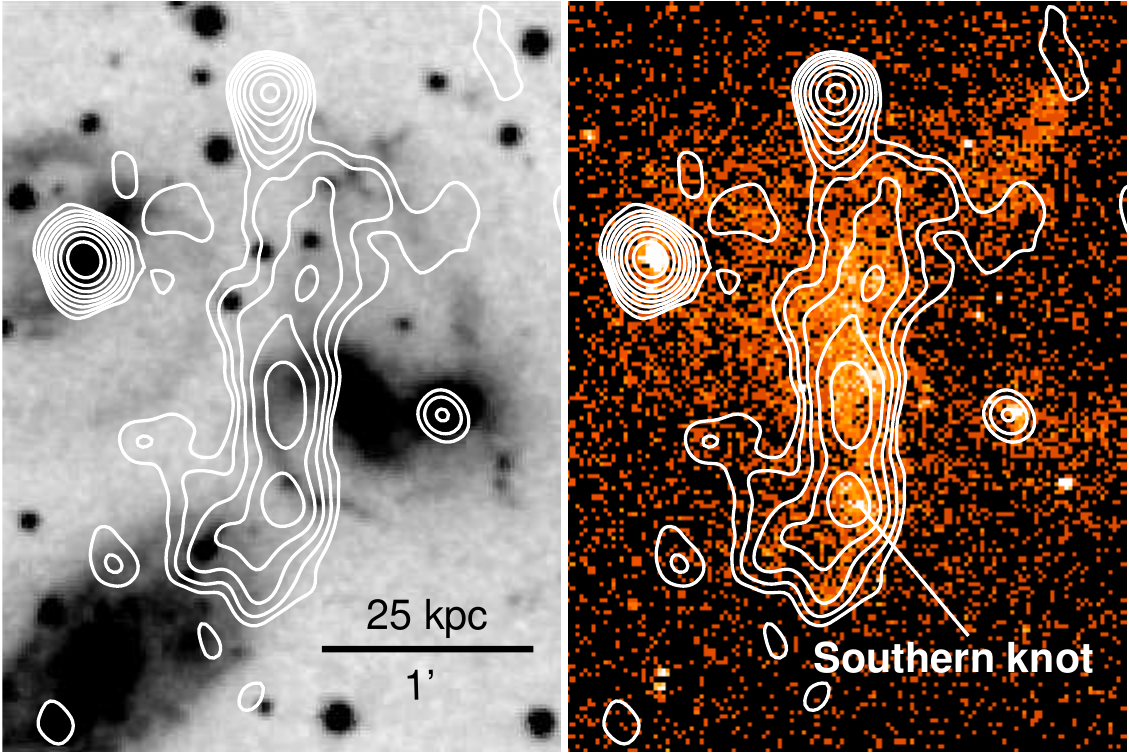}}
\caption{\label{GMRT} optical and 0.3-2.0~keV \chandra\ X-ray images of SQ
  with Giant Metrewave Radio Telescope 610~MHz radio contours
  overlaid. Contour levels are spaced by factors of 2, the lowest level
  indicating a flux of 0.25 mJy beam$^{-1}$, corresponding to a 3$\sigma$
  significance. The beam size is 6$\times$5\arcs.}
\end{figure}

Figure~\ref{GMRT} shows the radio continuum emission from the ridge,
measured at 610~MHz using the Giant Metrewave Radio Telescope. A detailed
analysis of these data will be presented in a later paper (Giacintucci et
al., in prep), but the contours confirm the general structure seen at
1.4~GHz \citep[e.g.,][]{AllenHartsuiker72,Williamsetal02,Xuetal03}. The
radio and X--ray emission generally agree, with most emission arising from
the ridge or the active nuclei of the galaxies. The brightest area of radio
emission associated with the ridge is the southern knot which is also a
region of bright X--ray emission, next to two bright optical and UV clumps
in the southern spiral arm of NGC~7318b. Outside the southern knot, the
radio and X--ray peaks are more poorly matched. The brightest X--ray
emission is found in the northern part of the ridge, whereas the radio is
brighter to the south, overlapping with the X-ray where the spiral arm
turns south, to the east of the core of NGC~7318b. This again is a UV
bright region, so it appears that the areas of strongest radio emission in
the ridge may be spatially correlated with star--forming regions. The
610~MHz radio emission also extends into the ``starburst A'' star forming
region (see Figure~\ref{images}) identified by \citet{Xuetal99}, where
X--ray emission is faint, but is anti-correlated with the X--ray clump to
the northwest of that region.

\section{Spectral Analysis}

\begin{figure}
\centerline{\includegraphics[width=8cm]{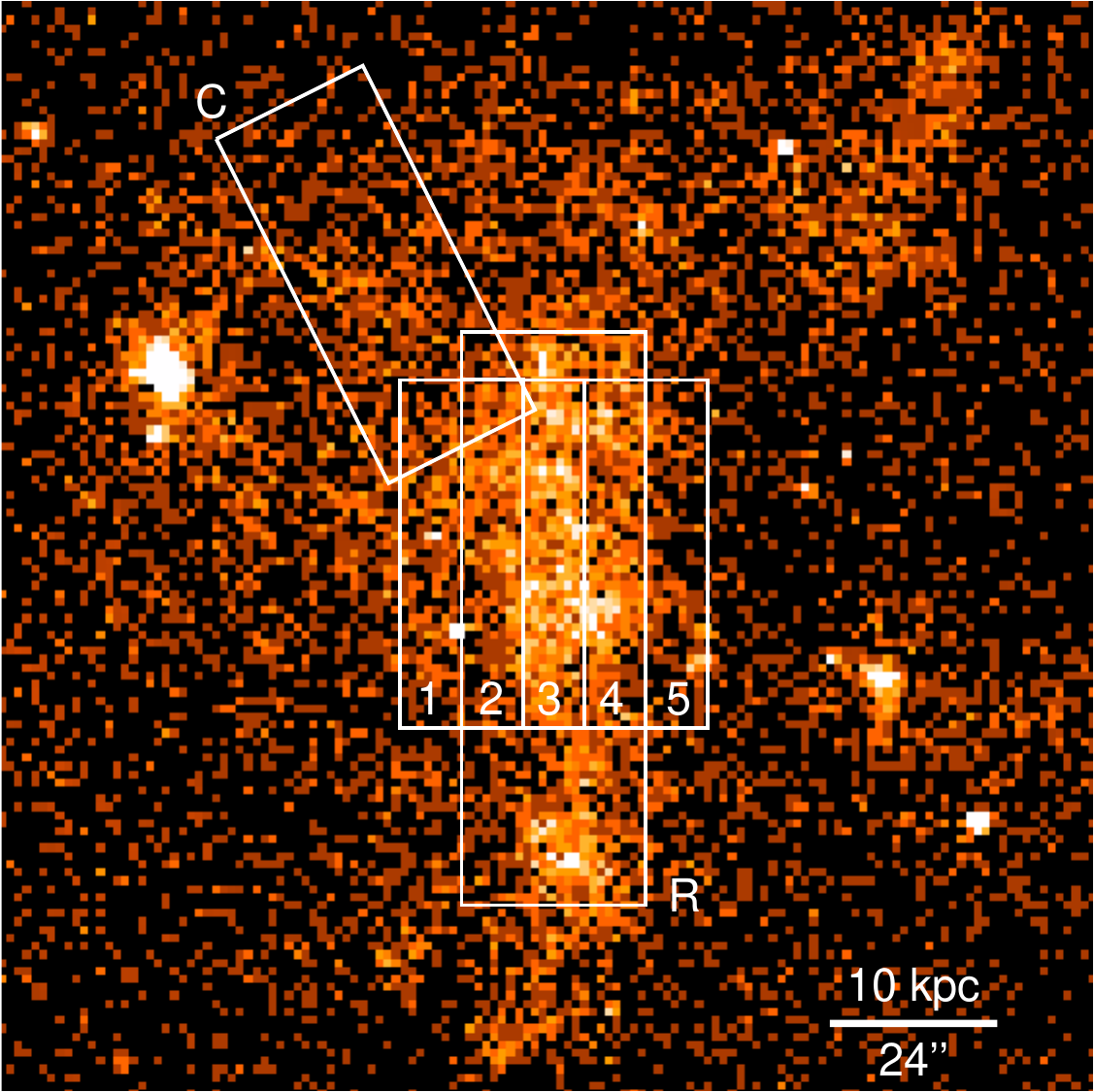}}
\caption{\label{specreg} \chandra\ 0.3-2.0 keV exposure corrected combined
  image from both exposures, binned by a factor of 2. Regions used to
  extract spectra and referred to in the text are marked.}
\end{figure}

In order to determine the properties of the gas in the ridge, and test the
accuracy of the temperature map, we extracted spectra from a number of
regions, shown in Figure~\ref{specreg}. Region R is a $77\times22$\arcs\
rectangle which covers the entire ridge, and region C, $46\times20$\arcs,
is chosen to match the coolest region of the temperature map. For region R,
fitting an APEC model with galactic absorption we find a temperature of
0.609$^{+0.024}_{-0.023}$~keV and abundance of
0.266$^{+0.076}_{-0.051}$\Zsol, agreeing well with the temperature map.
There is no evidence of excess absorption; freeing the hydrogen column does
not improve the quality of the fit, and does not significantly alter the
temperature or abundance.  The parameters also agree with those found by
\citet{Trinchierietal03} for an equivalent region in the earlier \chandra\
observation, and the central part of the ridge from the \xmm\ observation
\citep{Trinchierietal05}.  However, while the fit is statistically
acceptable ($\chi^2=68.94$ for 70 degrees of freedom), the spectrum shows
some evidence of an excess above 2~keV. \citet{Trinchierietal05}, fitting
\xmm\ spectra for the ridge as a whole, found that a second component was
required. For region C, we find that a powerlaw component is required in
addition to the plasma model, probably arising from X-ray binaries in the
spiral arms of NGC~7319, which the region overlaps. The gas temperature,
0.490$^{+0.060}_{-0.055}$~keV again agrees with the temperature map and is
significantly cooler than the ridge emission. The abundance is poorly
constrained (0.273$^{+0.368}_{-0.134}$\Zsol).

\begin{figure}
\centerline{\includegraphics[width=8cm]{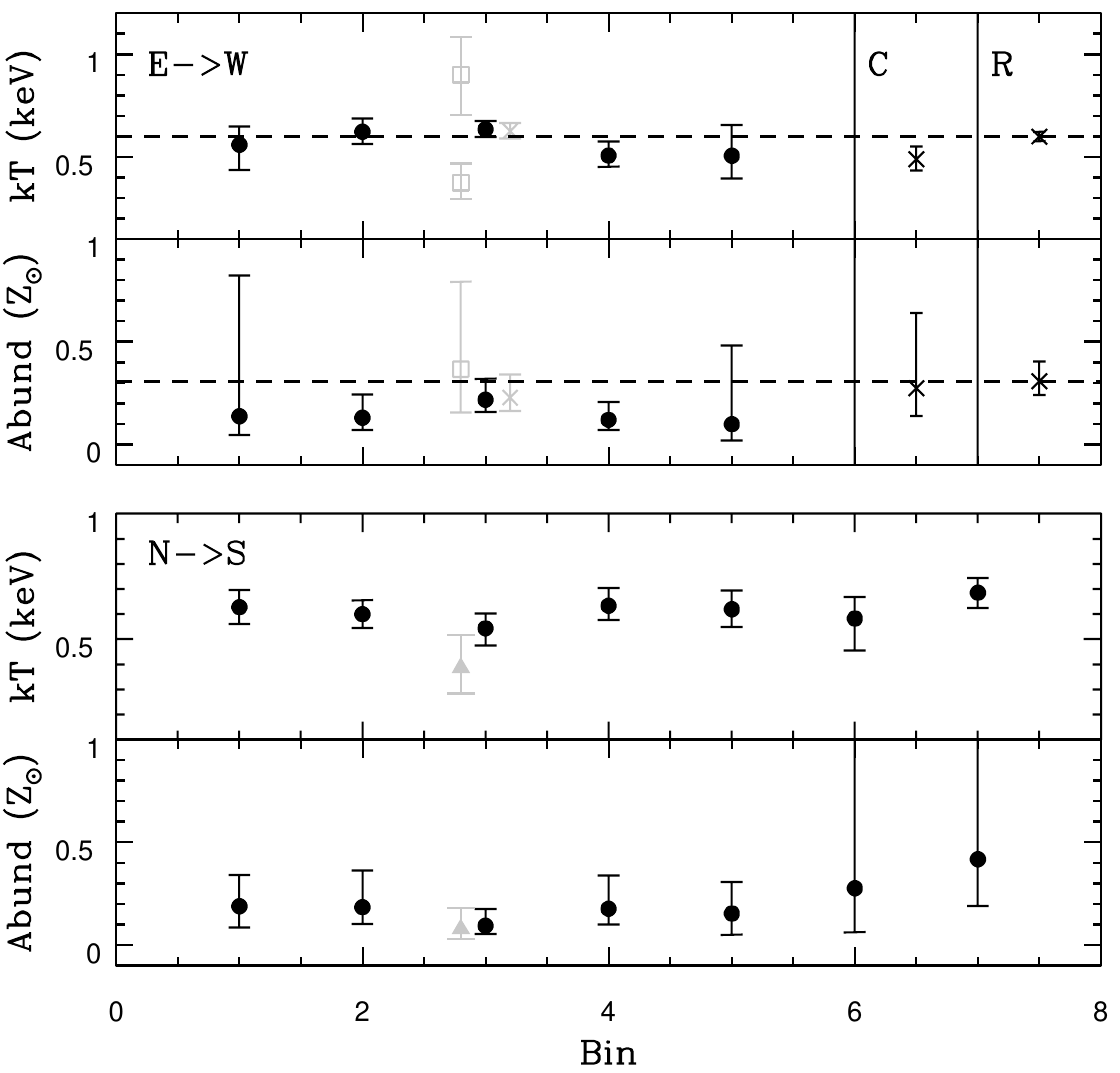}}
\caption{\label{TZ} Temperatures and abundances derived from spectral fits
  to the deep \chandra\ data. The upper panel (E-$>$W) shows results from
  five strips running across the ridge (regions 1-5 in
  Figure~\ref{specreg}). For comparison, best fit results from regions R and
  C are shown on the upper right. Dashed lines mark the best fit
  temperature and abundance in region R. The lower panels show the result
  of splitting region R into 7 bins to produce a profile running north to
  south (N-$>$S). Filled circles indicate APEC model fits with hydrogen
  column frozen at the galactic value, triangles APEC models with fitted
  column, crosses APEC+powerlaw models, and squares two-temperature APEC
  models. Note that for region R, the points represent an APEC+powerlaw fit
  with power law parameters frozen at the values expected for the HMXB
  population.}
\end{figure}

To search for any spatial variations in properties in the ridge, we derived
spectral profiles across and along the region. Figure~\ref{specreg} shows
five narrow rectangular regions spaced across the brightest part of the
ridge. Fitting spectra extracted from these with absorbed APEC models, we
find the temperatures and abundances shown in the upper panel of
Figure~\ref{TZ}. The only statistically significant difference is between
the temperatures in bins 3 and 4, which becomes less than 90\% significant
if hydrogen column is allowed to vary.  Bin 3 shows some evidence of an
excess of hard emission, which can be modelled by either a power law or by
using a two-temperature plasma model; either provides a statistically
acceptable fit. The remaining bins have fewer counts (200-400, compared to
the $\sim$870 net counts in region 3) so we cannot be certain that the hard
excess arises only from this region. To measure variation along the length
of the ridge, we split region R into seven equal bins. Results of absorbed
APEC model fits to these regions is shown in the lower panels of
figure~\ref{specreg}. We find consistent temperatures and abundances along
the length of the ridge, the only variation being the poorly constrained
abundances in bins 6 and 7, which cover the southern knot and the fainter
region immediately north of it. One notable feature is that the fit for bin
3 is significantly improved if hydrogen column is allowed to vary, in which
case its temperature is significantly cooler than the other bins. This is
the only region for which we find evidence of excess absorption, with a
best fit column of 20.1$^{+16.7}_{-11.0}\times10^{20}$\pcmsq, compared to
the galactic column of $6.17\times10^{20}$\pcmsq.

The ridge overlaps an extended region of star formation in the southeast
spiral arm of NGC~7318b and in the arm--like feature extending north to
Starburst A, and we therefore expect a hard contribution from High-Mass
X-ray Binaries (HMXBs). The star forming regions can be traced in the
ultraviolet (see Figure~\ref{images}) and \citet{Xuetal05} estimate the
star formation rate (SFR) in various parts of SQ based on the GALEX
extinction corrected FUV magnitude. Our region R overlaps the most of their
region V, so we adopt the SFR for this region and convert it to an expected
0.2-10~keV flux from high mass X-ray binaries (HMXBs) using the relation of
\citet{Grimmetal03}.  We assume a power law spectrum with $\Gamma=1.8$ for
the HMXBs, with normalisation fixed to produce the expected flux. The
resulting fit is somewhat under-constrained (red. $\chi^2_\nu$=0.845 for 71
d.o.f.) but the temperature and abundance (shown in Figure~\ref{TZ}) are
similar, within the uncertainties, to those measured previously. The power
law provides a good match to the expected hard component and if the
normalisation is freed, the predicted flux is within the 90\% uncertainty
bounds. We therefore conclude that the hard emission observed in the ridge
probably arises from HMXBs rather than a higher temperature gas
component. Although there are insufficient counts in region 3 to
definitely characterise the hard excess as a powerlaw, it seems reasonable
to assume that it also arises from HMXB emission.

To estimate the mean gas density and total gas mass in the ridge we model
the ridge as a cylinder described by region R ($\sim4.5$~kpc radius,
$\sim$28.5~kpc length). We can measure the gas mass in the brightest (and
as temperature is roughly constant, densest) part of the ridge using region
3 of the profile taken across the ridge, approximating the contribution
from foreground and background emission by using region 2 as a background
and again assuming a cylindrical distribution. While there is clearly
variation in density within these regions, these estimates provide a
measure of the mean density and a basis from which to examine the origins
of the gas. We include a powerlaw component to represent the HMXB
population and found spectral properties very similar to those described
above. The gas density in the ridge was found to be
1.167$\times10^{-2}$\pcmcu\ and the total mass
5.35$^{+2.39}_{-2.36}\times10^8$\Msol. The peak density is
2.34$\times10^{-2}$\pcmcu, corresponding to a mass of
$\sim7.25\times10^7$\Msol. This is similar to the density found by
\citet{Trinchierietal03}.

\begin{figure}
\centerline{\includegraphics[width=8cm]{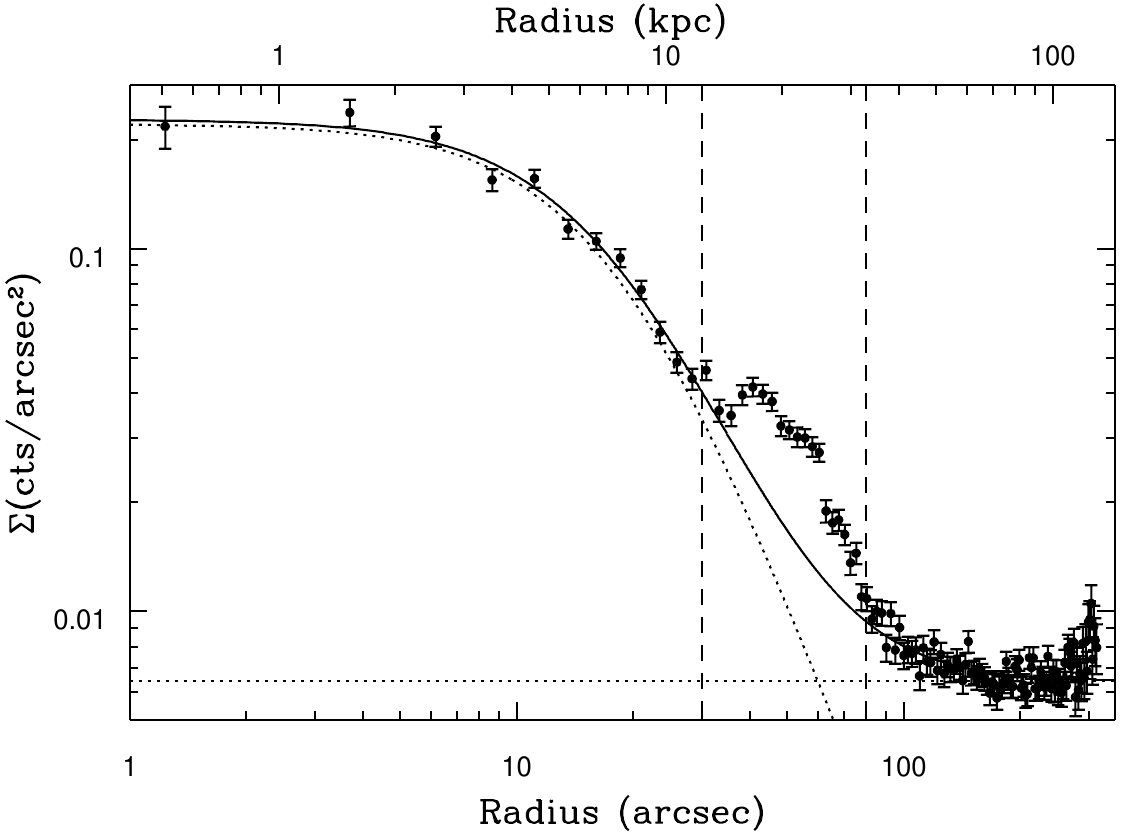}}
\caption{\label{SB} 0.3-2.0~keV exposure corrected radial surface brightness profile of
  SQ. Dotted lines show model components and the solid line the best fit
  model. The region bordered by dashed lines was excluded from the fit.}
\end{figure}

To estimate the mass of gas in the group halo, we first fit a surface
brightness profile in the 0.3-2.0~keV band, where gas emission should
dominate (Figure~\ref{SB}). We center the profile at the midpoint of the
northern section of the ridge, roughly on a line between the nuclei of
NGC~7319 and NGC~7318a.  We exclude regions corresponding to the cores of
the galaxies as well as a wedge chosen to remove the tail of emission
extending southeast \citep{Trinchierietal03}. As expected for such a
disturbed system, the profile is not well described by simple beta model,
but if an annulus between 30\arcs\ and 80\arcs\ is excluded a reasonable
fit is achieved. The resulting model underestimates the surface brightness
in the excluded region, which contains most of the lower luminosity diffuse
structures around the galaxies and ridge. The best fitting parameters are
R$_{\rm core}=19.58_{-1.27}^{+1.39}$\arcs\ and
$\beta=0.689_{-0.023}^{+0.026}$.  From these parameters and the peak
density we can determine the density profile of the gas, and numerical
integration produces a gas mas of 2.84$^{+1.45}_{-0.79}\times10^{10}$\Msol\
within 200\arcs\ ($\sim82$~kpc), approximately 50 times the mass of gas in
the ridge. Within the central 35~kpc (85\arcs), the approximate limit of
diffuse structures associated with the galaxies, we find a gas mass of
1.04$^{+0.49}_{-0.27}\times10^{10}$\Msol.  There are clearly considerable
uncertainties in these values, as we do not know the true distribution of
gas in the group, but they provide a useful estimate.

\section{Discussion} 
\label{sec:discuss}
There are three potential sources for the X--ray gas in SQ: accretion and
shock heating of primordial gas; star formation and the associated stellar
winds and supernovae; and shock heating of cold gas by the infall of NGC~7318b.
The latter suggestion has received the most support in the literature,
owing to evidence of shocks at other wavelengths
\citep{Shostaketal84,Xuetal03,Appletonetal06}.  The presence of optical and
\Hi\ tidal tails has lead to the suggestion that the hot gas in the group
core and particularly in the bright ridge is the product of shock--heating
of an \Hi\ filament created by previous tidal interactions between the
group and NGC~7320c \citep{Sulenticetal01,Xuetal03}. SQ contains
$\sim10^{10}$\Msol of \Hi\ \citep{Williamsetal02} but is \Hi\ deficient,
compared to predictions, by a factor $\sim$3
\citep{Verdes-montenegroetal01}. Our estimated total X--ray gas mass would
be sufficient to make up this lack, again suggesting that the lost \Hi\ has
been heated prior to the current interaction. However, given the
conflicting indicators from various observations, we consider all three
possibilities.

\subsection{Accretion of primordial gas}
We would expect the temperature of accreted primordial gas to correspond to
the Virial temperature of the system, which can be estimated from the
velocity dispersion of the group galaxies. Unfortunately the small number
of galaxies and uncertainty over which galaxies are actually
gravitationally bound make this a difficult quantity to estimate.
\citet{OsmondPonman04}, using an iterative technique to determine group
members and including a correction for the effects of biasing in systems
with small numbers of galaxies, find a velocity dispersion
$\sigma_v=467\pm176$\kmps for five galaxies (NGC~7317, 7318a, 7318b, 7319,
7320c), equivalent to a virial temperature T$_{\rm vir}\sim1.3$~keV. If, as
the observations suggest, NGC~7318b is passing through the group for the
first time, it should not be included in the estimate. We therefore
recalculate $\sigma_v$ using the same methods as \citet{OsmondPonman04},

\begin{equation} 
\sigma_v = \sqrt{\frac{\Sigma(v-\overline{v})}{N-\frac{3}{2}}} \pm
\frac{\sigma_v}{\sqrt{2(N-\frac{3}{2})}} \kmps,
\end{equation}

where $v$ is the recession velocity of the individual galaxy,
$\overline{v}$ is the mean recession velocity of the group and $N$ is the
number of galaxies. We adopt velocities NGC~7317, NGC~7138a and NGC~7319
from \citet{Molesetal97}, and take the velocity of NGC~7320c from the
HyperLeda database\footnote{http://leda.univ-lyon1.fr/}. We find
$\sigma_v=345\pm154$\kmps, equivalent to T$_{\rm vir}\sim0.7$~keV. This is
somewhat hotter than the temperatures seen in the group core, but is in
reasonable agreement with the temperature map at larger radii. \xmm\
imaging shows that SQ possesses a moderately extended (130-150~kpc) diffuse
halo, whose extent strongly argues that it must predate the current
interaction \citep{Trinchierietal05}. A spectrum extracted from an
85-135\arcs\ annulus confirms the higher temperatures seen in the map, with
kT=0.83$^{+0.12}_{-0.09}$~keV. This agrees reasonably well with the
expected T$_{\rm vir}$ for four galaxies. However, we note that this value
depends primarily on the recession velocity of NGC~7320c, since the other
three galaxies have almost identical velocities. The inherent uncertainties
in such a measurement therefore make this result inconclusive.

\subsection{Star formation and Supernovae}

A possible source for some of the emission in the group core and ridge is
star formation and the resulting supernovae. Our spectral fitting results
suggest that the hard component is consistent with emission from HMXBs in
the young stellar populations of the star forming regions of NGC~7318b
which the ridge overlaps. It is very noticeable that many of the brightest
regions of the X--ray and radio emission in the ridge correspond to
UV--bright knots in NGC~7318b, which are likely to be sites of star
formation either now or in the recent past. The best example is the
southern knot marked in Figure~\ref{GMRT}, which lies between and just
outside two of the brightest UV regions in the southern part of NGC~7318b.
This region is also associated with \Ha\ emission at the velocity of
NGC~7318b \citep{Sulenticetal01} and hosts emission line features which
indicate the presence of shocks \citep{Xuetal03}. It therefore seems likely
that both star formation and shock heating may be involved in the X--ray
emission of this region, and perhaps the ridge as a whole. It is notable
that if we assume simple radiative cooling and fit the spectrum for region
R with a cooling flow model, we find a mass deposition rate of 1-4\Msol\
yr$^{-1}$, quite consistent with the star formation rate of
$\sim$1.5\Msol/yr \citep{Xuetal05}.

The starburst A region is a notable counterexample. Its SFR is similar to
that of the ridge as a whole \citep{Xuetal05}, but its X--ray emission is
considerably fainter. In principle this may be in part due to absorption by
the $\sim3\times10^9$\Msol\ of \Hi\ associated with the region
\citep{Williamsetal02}. X--ray spectral fits for this region show no
evidence of excess absorption, but the number of detected counts is low and
the uncertainties on measured \nh\ necessarily large. Similarly, the X--ray
and radio emission to the southeast of the southern knot, extending toward
NGC~7320, cannot be associated with star formation, suggesting shock
heating must be important there. Accurate abundance measurements on small
scales might help resolve this issue by allowing us to identify regions
where rapid star formation has enriched the X--ray gas, but unfortunately
even our deep \chandra\ observation contains too few counts for such
detailed mapping.

We can also estimate the soft emission likely to arise from star formation
activity. \citet{Ranallietal03} propose a correlation between SFR and
0.5-2~keV X-ray emission, in the absence of uncorrected absorption. Based
on this correlation and the FUV SFR, we would expect a luminosity
L$_{\rm0.5-2~keV}=6.8\times10^{39}$\ergps, a factor $\sim$8.5 less than the
observed gas luminosity in region R,
L$_{\rm0.5-2~keV}=7.6\times10^{40}$\ergps. However, comparisons of the
X-ray and UV SFRs of star-forming galaxies in the \chandra\ deep field
south show a scatter of a factor 10 or greater between the two wavebands
\citep{Rosa-Gonzalezetal07}. It therefore seems likely, but not certain,
that some other source of soft emission beyond the star formation in the
region is required.

Another approach is to consider the energy available from the stellar
population. A simple approximation would be to assume that one supernova
will be produced per 100\Msol\ of stars formed, and that each supernova
yields approximately 10$^{51}$ erg, in which case we would expect an energy
injection rate in the ridge of $\sim5\times10^{41}$\ergps. However, the
lifespan of stars capable of producing supernovae is comparable to that of
the collision ($\sim4\times10^7$~yr compared to 2-8$\times10^7$~yr), so it
is likely that a significant fraction of the expected supernovae have yet
to occur, if the star formation was triggered or enhanced by the infall of
NGC~7318b into SQ. An alternative method is to estimate the supernova rate
from the FIR luminosity of the ridge using the relation of
\citet{Cappellaroetal99}. The available 60~$\mu$m and 100~$\mu$m fluxes are
likely to be overestimates, since they are derived from low resolution
\textit {Infrared Space Observatory} (ISO) observations and include
emission from NGC~7318a/b \citep{Xuetal03}. Based on these fluxes, we find
an expected SN rate of 4.9$\times10^{-3}$~yr$^{-1}$.  We would thus expect
$\sim1.5\times10^{41}$\ergps\ of energy to be injected into the gas in the
ridge. These values are comparable to the reported IR and H$_2$
luminosities \citep[2.5$\times10^{41}$\ergps\ and 8.4$\times10^{40}$\ergps\
respectively,][]{Appletonetal06}, but a significantly larger value would be
required to explain all emission from the ridge.

Supernovae will also inject metals into the gas in the ridge. Assuming only
type II supernovae and an elemental yield of $\sim$0.09\Msol\ SN$^{-1}$ of iron
\citep{Nomotoetal97a}, we find an expected abundance in the ridge of
0.267\Zsol, in good agreement with measured values. However, this is rather
surprising given the multi--phase nature of the gas in the ridge; we might
have expected the enriched material to be poorly mixed, and for
multi--temperature emission to lead us to underestimate the abundance. In
general, it seems clear that while star--formation and supernovae play a
role in producing the ridge emission, they are unlikely to be the dominant
source of energy.

\subsection{Shock heating}

The remaining alternative is that the infall of NGC~7318b into the group
core has caused shock heating in a pre-existing intergalactic medium.
Several of our X--ray results support this hypothesis and indeed it is
difficult to imagine circumstances in which the interaction would not
produce shocks. The low measured abundances ($\sim$0.3\Zsol or less) would
be expected for gas originally in the extended \Hi\ disks in the outskirts
of spiral galaxies. Our estimated total mass of X-ray emitting gas in the
central 35~kpc, $\sim1\times10^{10}$\Msol, is comparable with the masses of
\Hi\ in the tidal tail ($\sim4\times10^9$\Msol) and starburst A regions
\citep[$\sim3\times10^9$\Msol,][]{Williamsetal02}, which would have been
linked by the now--heated \Hi\ filament. The lack of excess absorption
along the filament indicates that there is little \Hi\ now in the region,
suggesting it has been efficiently shocked or removed by some other means.

However, the lack of evidence for a significantly hotter gas component is
problematic. We can estimate the post-shock temperature of the gas based on
some simple assumptions about its pre-shock state. We consider three
scenarios: 1) the pre-shock gas was \Hi, 2) the $\sim$0.4~keV component of
the multi-temperate fit to region 3 represents the pre-shock state, or 3)
the 0.6~keV gas represents the pre-shock material. We assume a shock
velocity of 850-900\kmps, based on the relative velocity of NGC~7318b with
respect to the three galaxies in the group kernel. While the inclusion of
NGC~7320c would lower the velocity difference to 700\kmps\ we note that the
higher value is supported by estimates from \Hi\ velocities
\citep{Williamsetal02} and emission line studies
\citep{Appletonetal06,Ohyamaetal98}.

For scenario 1, this would yield a strong shock, heating the \Hi\ to
$\sim$1.2~keV. Scenarios 2 \& 3 produce weaker shocks with mach numbers
$\mathcal{M}=2.3-2.9$ and expected post-shock temperatures of 1.3-1.5~keV.
Our spectral fits suggest that all hard emission in the ridge is likely to
arise from HMXBs, ruling out any such high-temperature component. Even if
we assume that HMXB emission is unimportant and instead use an additional
plasma model, the upper limit on the temperature is lower than that
predicted for shock heating, and only a small fraction of the gas would be
in the higher temperature component, very unlikely if the shock is ongoing.

Several possible explanations of these lower temperatures can be suggested:
1) the shock might be oblique rather than perpendicular, 2) the gas has
cooled through adiabatic expansion, 3) rapid gas cooling via collisional
heating and sputtering of dust would allow the temperature to drop swiftly
after the shock passed, or 4) the shocked gas is not in ionisation
equilibrium, leading us to underestimate its temperature.
 
Adiabatic expansion of the gas would explain both its cool temperature and
the small mass of X-ray gas in the ridge compared to that expected in a
pre-existing \Hi\ filament. We can estimate an expansion timescale from the
speed of sound in the gas, $\sim$390\kmps for the gas in the ridge and
$\sim$310\kmps for the cooler material in region C. At such velocities, the
sound crossing time for a distance of 35~kpc (from the ridge to the edge of
region H) is 88~Myr. This is similar to the estimated total timescale of
the encounter between NGC~7318 and SQ \citep[20-80~Myr,][]{Sulenticetal01}
and the synchrotron decay timescale of the radio emission associated with
the ridge \citep[80~Myr,][]{VanderhulstRots81}, and longer than the
estimated ages of the stellar populations in Starburst A
\citep[10-20~Myr,][]{Xuetal99} and the bluest star clusters in SQ
\citep[5-7~Myr,][]{Sulenticetal01}. It is also notable that the temperature
map shows a temperature rise with distance from the ridge, rather than the
decline that would be expected.  As the interaction is thought to be
ongoing, this strongly suggests that adiabatic expansion cannot have
greatly cooled the gas, and that the X-ray emitting gas we observe cannot
arise entirely from shock--heating of a pre--existing \Hi\ filament. This
suggests either that some other source of gas has created the large scale
X--ray halo, or that some prior interaction caused heating of part of the
\Hi\ in the group at an earlier epoch. The \Hi\ distribution in NGC~7318b
has been interpreted as supporting a past interaction with SQ
\citep{Williamsetal02}, and the presence of X--ray emission coincident with
the tidal tails \citep{Trinchierietal05} provides further support.

\citet{Xuetal03} suggest the possibility that the shocked X--ray gas is
efficiently cooled on short timescales by collisional heating of dust. They
estimate a cooling timescale for this mechanism of only
$\sim2.1\times10^6$~yr and find a close match between the expected and
observed FIR luminosities in the shock region. They also point out that the
FIR luminosity is about an order of magnitude greater than the X-ray
luminosity, supporting dust as the dominant source of energy
loss. Comparison of \textit{Spitzer} Space Telescope 160~$\mu$m images with
the \xmms\ data confirms the presence of FIR emission in the general area
of the X-ray/radio ridge \citep{Xuetal08}. However, this scenario may not
be able to explain the consistent $\sim$0.6~keV temperatures across the
shock. The dust cooling timescale is so short that we would expect the
shock to move only a few arcseconds, but we would expect to see a strong
temperature gradient across this region, from high temperature gas at the
shock front to very cool gas further downstream. There is no reason to
expect cooling to stop at the observed temperature, so a coincidence would
be required for us to find the gas in the ridge at the same temperature as
its surroundings. An initial comparison between the \textit{Spitzer}
160~$\mu$m data and \chandra\ imaging suggests that X-ray ridge is offset
slightly to the west of the FIR emission, most notably in the southern part
of the ridge. Unfortunately the lower resolution of the \textit{Spitzer}
image and the presence of bright sources associated with star-forming
regions, NGC~7319 and NGC~7320 make detailed examination of any offset
difficult.

An alternative view is presented by \citep{Guillardetal09} who suggest that
variations in the pre-shocked material will lead to a multi--phase
post--shock environment. The least dense pre--shock gas will be heated to
the highest temperatures and although collisional heating of dust will
initially be the most efficient cooling mechanism, the dust will be rapidly
destroyed by sputtering before the gas cools by more than $\sim$0.1~keV.
Denser pre-shock regions will experience slower shock velocities, be heated
to lower temperatures, and rapidly cool to form clouds of molecular gas.
This scenario has the advantage of explaining the high H$_2$ luminosity of
the ridge \citep{Appletonetal06} while not requiring a coincidence of
timing to produce the observed temperature distribution. Taking into
account the initial cooling by dust grains, Guillard et al. find that the
observed temperature of 0.6~keV would be produced by shock of velocity
800\kmps, similar to the observed velocity difference. The scenario does
not explain why the ridge has such a similar temperature to its
surroundings, and it is unclear what relationship should be expected
between the X-ray and FIR emission; dust associated with gas which is being
shocked now should be highly luminous, but if most of the dust is
associated with the \Hi\ filament then we might expect the FIR morphology
to follow that structure.

\begin{figure}
\centerline{\includegraphics[width=8cm]{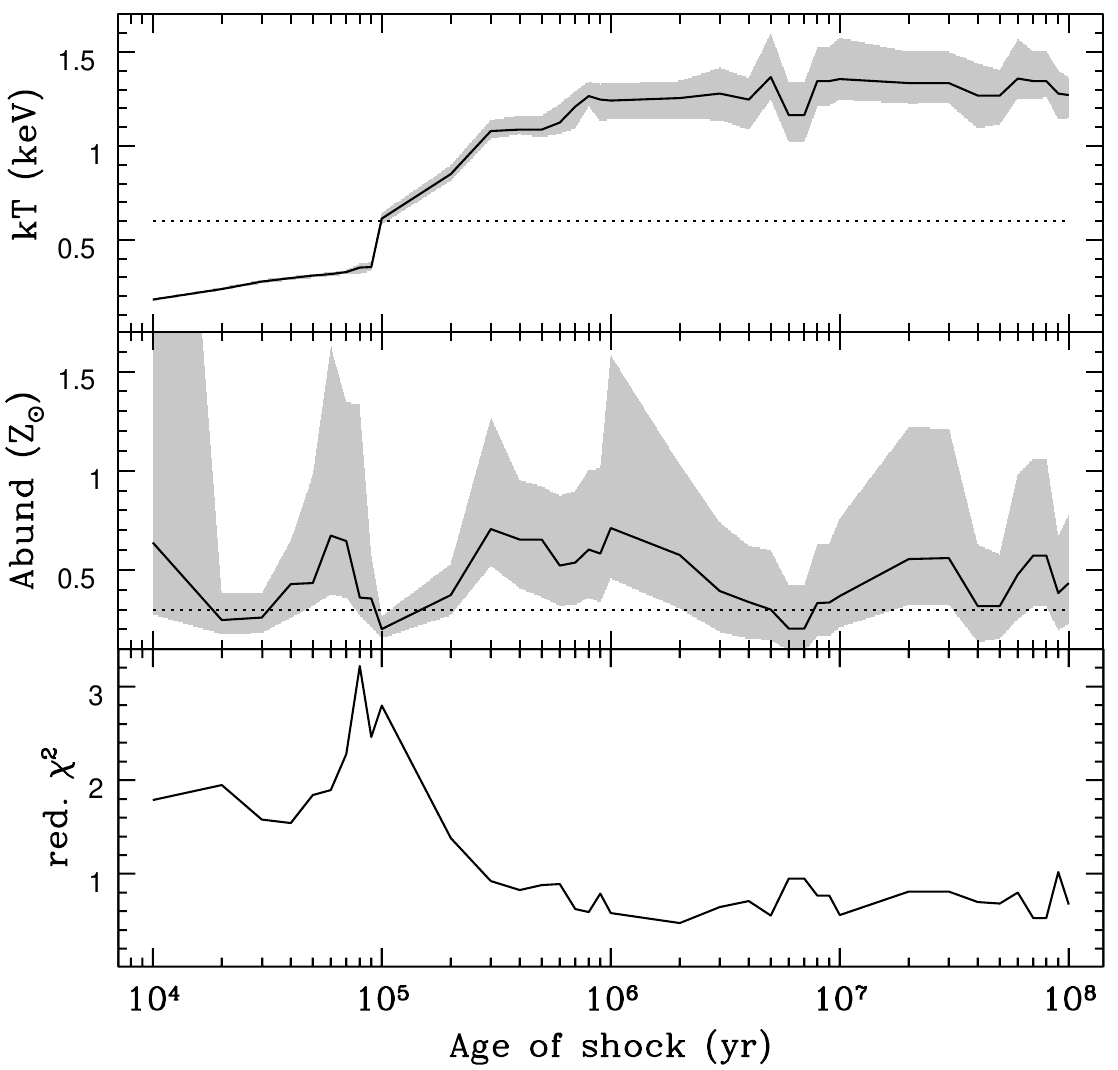}}
\caption{\label{NEI} Results of fitting simulated non-equilibrium
  ionisation (NEI) models of various ages with simple APEC models. Solid
  lines mark the best fit values, grey regions the 90\% uncertainties. The
  simulated temperature and abundance were 1.35~keV and 0.3\Zsol\
  respectively.  The APEC model fit accurately measures temperature for
  ages older than $10^6$~yr. The observed temperature and abundance for
  region R are marked with dotted lines. NEI models which find the observed
  temperature and abundance have very high reduced $\chi^2_\nu$ values,
  demonstrating obvious differences in spectral shape from a simple
  equilibrium plasma.}
\end{figure}

To test the possibility that the gas might be out of ionisation
equilibrium, we followed the example of \citet{Grahametal08} and simulated
the expected spectrum of the gas at various times after the shock, assuming
a post--shock density equal to that measured in the ridge, a true
post-shock temperature of 1.35~keV and an abundance of 0.3\Zsol. We also
assume the density determined for region R, 1.167$\times10^{-2}$\pcmcu, but
note that the model uses a parameter $\tau=n_e*age$, so increasing density
by a factor of 2 simply shifts our results by a factor of 2 in age.  We
included a powerlaw component to account for HMXBs, and fitted the
resulting spectra with APEC+powerlaw models. We find that after $\sim$1~Myr
the fitted temperature is an accurate measure of the true temperature of
the gas. To produce the temperature and abundance we observe, a post-shock
age of only $\sim10^5$~yr is required, and at this stage the spectrum
clearly deviates from that expected for an equilibrium plasma (reduced
$\chi^2_\nu\sim2$). It therefore seems unlikely that a significant fraction of
the gas is out of ionisation equilibrium.

The possibility that the shock is largely oblique is perhaps more
promising. If NGC~7318b is moving primarily along the line of sight with
the eastern or north--eastern edge of its disk leading
\citep{Sulenticetal01}, then some degree of obliquity would be expected for
all parts of the galaxy except the leading edge. The region producing a
perpendicular shock might be relatively small, producing only a minor
contribution to the gas temperature along the line of sight, with other
regions, including the section of the disk responsible for shocking the
ridge, producing less energetic heating.  Using our assumed velocity of
900\kmps\ and post--shock temperature 0.6~keV, we find that for a strong
shock (i.e., if the pre--shock material is \Hi) an angle of 34\degree\
would be required (as compared to 90\degree\ for a perpendicular shock), in
close agreement with the estimate of \citet{Trinchierietal03}. For weak
shocks, obliquity is clearly unhelpful if the pre-shock gas had a
temperature of 0.6~keV, but for 0.4~keV material a similar angle,
$\sim$31\degree\ would produce a post--shock temperature of 0.6~keV. These
angles are comparable to the angle of the disk of NGC~7318b to the line of
sight, 30-40\degree, estimated from optical imaging \citep{Sulenticetal01}.

However, it is worth noting two potential problems with this scenario.
Firstly, the angle of the shock will be larger than the angle of the galaxy
to the line of sight, owing to the opening angle of the shock cone. For a
strong shock this angle will be narrow, but for a weak shock it is likely
to be quite wide, $\sim$70\degree\ for the $\mathcal{M}=2.9$ shock expected
in 0.4~keV gas. Secondly, while the interaction which has produced the ridge
may have involved \Hi\ or 0.4~keV gas, NGC~7318b must have passed through
higher temperature as on the far side of SQ as it fell into the core, and
should have shocked this material. While the angle of infall is very
unfavourable for detecting this shocked material, it seems surprising that
we do not see some evidence of hotter gas around NGC~7318b.

We can estimate the thermal energy available from a shock if it does heat
pre-existing \Hi\ to 0.6~keV. If we conservatively assume that a region
approximately the size of the brightest section of the ridge is directly
affected by the shock ($\sim25\times3$~kpc), and assume the peak density to be
representative of the post--shock gas, we expect
$\sim1.65\times10^{42}$\ergps to be injected into the gas. For an area
equal to the whole of region R with the average density determined for that
region, about twice this energy would be available. The energy available
from a shock is thus about an order of magnitude greater than that
available from star formation and supernovae. While we do not fully
understand how such a shock can produce the observed conditions, it seems
likely that shock--heating must play an important role in heating the gas
in SQ.

\section{Conclusions}

Stephan's Quintet provides a rare opportunity to observe a galaxy group in
the process of development from an X--ray faint spiral--rich system to a
more evolved, elliptical--dominated, X--ray bright state. The possibility
that a significant fraction of the cold gas in the group is being heated by
the ongoing dynamical interactions has an obvious importance for our
understanding of the origins of the hot gas halos of X-ray luminous groups.
We have analysed a deep, $\sim$95~ks \chandra\ observation of SQ with the
goal of placing strong constraints on the nature of the interactions in the
system and improving our understanding of the origin of the emission ridge.
Our results can be summarized as follows:

\begin{enumerate}

\item The X--ray emission in the ridge is well described by a plasma of
  temperature 0.6~keV and abundance 0.3\Zsol\ plasma, with a powerlaw
  component consistent with that expected from HMXBs associated with the
  star formation in this region. The fitted temperature and abundance are
  very similar to the values found for the diffuse emission surrounding the
  ridge. This strongly suggests that the gas cannot be the product of a
  normal shock driven by the interaction with NGC~7318b. An oblique shock
  with an angle of $\sim30$\degree\ could produce the temperatures we
  observe if the pre--shock material were \Hi\ or $\sim0.4$~keV gas. This
  angle matches that of the disk of NGC~7318b, suggesting that the shock
  heating of an \Hi\ filament is a viable mechanism to produce much of the
  X--ray gas observed in the emission ridge. However, for 0.4~keV or hotter
  gas, the opening angle of the shock would be wide, and the interactions
  with the surrounding, pre--existing X--ray halo must at some point
  produce a normal shock. It is therefore unclear why no high--temperature
  plasma component, associated with shocking of the pre--existing hot gas,
  is observed.

\item The total mass of X--ray emitting gas in SQ is similar to the deficit
  in \Hi\ compared to expected values. It is therefore possible that the
  hot gas component of the group halo arises entirely from shock heating of
  \Hi. However, the extent of X-ray emission is large compared to the
  expected rate of expansion of a shock heated filament of \Hi. This
  strongly suggests that a hot intra--group medium was already present in
  the group before the current interaction. This could have arisen during
  previous dynamical interactions, for which there is plentiful evidence,
  in particular multiple tidal tails pointing roughly toward the nearby
  galaxy NGC~7320c. Alternatively, the hot gas component could be
  primordial material shock heated during its accretion into the group. The
  expected Virial temperature of SQ, based on the group velocity dispersion
  excluding NGC~7318b, is similar to the observed gas temperature outside
  the group core, though with large uncertainties owing to the small number
  of galaxies involved and the dynamical state of the group.

\item Some of the brightest X-ray and radio emission regions in the ridge
  appear to be spatially correlated with star forming regions, identified
  by UV emission. The southern knot and the point at which the eastern
  spiral arm of NGC~7318b turns south are both bright in X--ray and radio.
  However, outside these regions there are significant disagreements
  between radio and X--ray distributions. This supports the idea that a
  combination of emission mechanisms is required to produced the structures
  observed.

\item The cooling rate of hot gas in the ridge is consistent with the
  current star formation rate, suggesting that star formation in this
  region may be largely driven by radiative cooling. The mass of metals
  expected from the predicted number of supernovae also provides a good
  match to the observed gas abundance. However the soft X--ray luminosity
  expected from supernovae and star formation is smaller than that observed
  by a factor $\sim10$ and the energy available from supernovae associated
  with star formation is also too small to support the emission observed at
  various wavelengths. The energy available from shock heating is probably
  sufficient to produce the observed luminosities.

\item Efficient energy loss from the hot gas via collisional heating of
  dust appears unlikely to be the dominant source of cooling at present,
  since it would be expected to introduce strong temperature gradients
  which are not observed in the temperature map. Assuming that an \Hi\
  filament has been shocked, the dust associated with this filament would
  now be in the emission ridge, and would most effectively cool the gas in
  the densest parts of the ridge. As the ridge has a similar (or slightly
  warmer) temperature to its surroundings, a coincidence would be required
  for us to observe the gas at just this point in its cooling. Dust may
  have been a more important factor in cooling the gas immediately after
  the passage of the shock, but must have been rapidly destroyed by
  sputtering.

\item It seems unlikely that our measured temperatures are being biased by
  the presence of non-equilibrium ionisation gas. Testing suggests that a
  non-equilibrium plasma capable of producing the temperature and abundance
  we observe would be very poorly described by an APEC model and would have
  to have been shocked within the last $\sim10^5$~yr. By $10^6$~yr after
  shocking we would expect spectral fitting to accurately reproduce the
  plasma temperature. We therefore conclude that our measured temperatures
  are correct and that the great majority of the gas is in ionisation
  equilibrium.

\end{enumerate}

In general, we conclude that the source of much of the X--ray gas in the
ridge, and the majority of the energy powering the multi--wavelength
emission from this region, arises from shocks driven by the infall of
NGC~7318b. However, star formation plays an important role in determining
the exact morphology of the ridge and is responsible for the majority of
the hard emission via HMXBs. Considering the ridge in isolation, it appears
possible that \Hi\ in a pre--existing filament and in the disk of NGC~7318b
has been shock heated by the interaction, with additional heating and
enrichment by supernovae and stellar winds. However, this scenario cannot
explain the larger X--ray halo of SQ, which may be evidence of
shock--heating during previous dynamical interactions.

\acknowledgments The authors would like to thank S. Immler and D.~J. Saikia
for providing access to the \swift\ and GMRT data, and P. Nulsen, A. Zezas
and T.J. Ponman for helpful discussion of various aspects of the paper.
Support for this work was provided by the National Aeronautics and Space
Administration through Chandra Award Number G07-8133X-R issued by the
Chandra X-ray Observatory Center, which is operated by the Smithsonian
Astrophysical Observatory for and on behalf of the National Aeronautics
Space Administration under contract NAS8-03060. This research has made use
of software provided by the Chandra X-ray Center (CXC) in the application
packages CIAO, ChIPS, and Sherpa, and of the NASA/IPAC Extragalactic
Database (NED) which is operated by the Jet Propulsion Laboratory,
California Institute of Technology, under contract with the National
Aeronautics and Space Administration.

{\it Facilities:} \facility{CXO (ACIS)}, \facility{GMRT}, \facility{Swift (UVOT)}

\bibliographystyle{apj}
\bibliography{../../paper}

\end{document}